\documentclass[]{aastex701}

\newcommand{\edit}[1]{\textcolor{black}{#1}}
\newcommand{\secondedit}[1]{\textcolor{black}{#1}}

\begin{document}

\title{Dust On, Dust Off: HST Observations of the Newly Dormant Jupiter Co-orbital\\Comet P/2023 V6 (PANSTARRS)}

\correspondingauthor{John W. Noonan}

\author[0000-0003-2152-6987]{John W. Noonan}
\affiliation{Department of Physics\\
Edmund C. Leach Science Center\\
Auburn University\\
Auburn, AL 36849, USA}
\email{noonan@auburn.edu}

\author[0000-0003-1008-7499]{Theodore Kareta}
\affiliation{Department of Astrophysics and Planetary Science\\
Villanova University\\
Villanova, PA, 19085, USA}
\email{theodore.kareta@villanova.edu}




\begin{abstract}
As the Centaurs of the Outer Solar System have become better studied, their relationship with the Jupiter Family Comets and their connection to the primordial protoplanetary disk has come under deeper scrutiny: which properties of the comets observed closer to the Sun are original, and which were modified significantly en route during their Centaur phase? The active Jupiter co-orbital comet P/2019 LD2 (ATLAS) attracted significant attention after its discovery when it was realized that the object would transition between a Centaur and JFC orbit in 2063, the first time this change can be observed and monitored in real time.   In this study, we present new ground-based and Hubble Space Telescope imaging of the single other Jupiter co-orbital comet, P/2023 V6 (PANSTARRS), throughout late 2024 and early 2025. V6 was not detected in ground-based imaging after early October, and deep imaging with HST taken in early December is consistent with a $m_V = 28.06 \pm 0.03$ point source at its expected location. This both implies that the object ceased activity sometime in late 2024, and is thus in a different thermal state than LD2 despite both having large active fractions, and that V6's diameter is approximately $D\sim340$ meters. This is likely the smallest object directly observed by HST beyond the Main Belt, and certainly the smallest Centaur yet discovered, highlighting the \edit{ability to detect active sub-km} objects beyond Jupiter with current surveys and \edit{potentially probing smaller sizes with LSST}.

\end{abstract}

\keywords{}


\section{Introduction}\label{sec:intro}
The reservoirs of active small bodies in the solar system are defined through their ongoing evolution, and thus it is of wide utility to investigate and understand how comets, asteroids, and Centaurs become active -- especially when they are in orbits, have compositions, or are at heliocentric distances where traditional activity would not be expected \citep{Jewitt2024}. \edit{Many} comets and Centaurs \edit{have displayed }vigorous mass loss far beyond where water and even carbon dioxide sublimate easily \citep{Jewitt2017}. \edit{However, the contribution to activity magnitude and frequency from} factors like recent orbital changes \citep{2018P&SS..158....6F,Lilly2024}, size \citep{2025cent.book...10K}, and bulk volatile abundance \citep{HarringtonPinto2022} \edit{are} not well understood. The timescales of orbital evolution range from decades to millenia, and thus opportunities to watch objects move between populations or experience substantial changes in their thermal environments to see how they respond \edit{are} quite rare. Furthermore, given the small number of known active Centaurs and the wide variety of sizes and orbits in the population, there are few cases for which one can make an `apples to apples' comparison of two objects which share many of these properties. Much of our knowledge of the Centaurs is thus built on single objects which we hope are representative, and which is clearly biased towards these largest and most active of the population.

\edit{One mechanism to address this} would be to study objects which have been temporarily trapped as co-orbitals or temporary captures of the \edit{giant planets}, and thus should be in relatively similar orbits and thermal environments for decades or longer at a time. The first active Jupiter co-orbital P/2019 LD2 (ATLAS, hereafter `LD2') was discovered very strongly active for its heliocentric distance \citep[][]{Steckloff2020,Kareta2021,Bolin2021} despite being in a relatively low-eccentricity orbit and thus not experiencing large temperature swings. The importance of understanding and contextualizing LD2's strong activity was underscored by the realization that LD2 would transition between a Centaur orbit and an inner Solar System Jupiter Family Comet (JFC) orbit in 2063 -- the first time that observers can watch an ecliptic comet \citep{Levison1997} make this transition, potentially offering insight into how all of the JFCs have been altered by their trip through the Outer Solar System before they were studied in depth closer to the Sun. Four years later, the second active Jupiter co-orbital, P/2023 V6 (PANSTARRS, hereafter `V6') was discovered, prompting the observational and dynamical investigation of \citet{2024ApJ...967L...5K} which showed it to be producing approximately fifteen times less dust given a nearly identical orbit, observing circumstances, and analysis techniques. Whether or not this difference was driven by differing evolutionary states (e.g., V6 was more depleted of hypervolatiles than LD2) or differing sizes (e.g., V6 was significantly smaller than LD2) was unclear. Whether or not LD2 was `typical', or at least close enough to serve as a case-study to understand the Centaur-JFC transition, rests on how to interpret its differences with V6 -- the only other known active co-orbital or temporary capture of a Giant Planet.

In this paper we present the results of our December 2024 observations of P/2023 V6 (PANSTARRS) using the WFC3 instrument on the Hubble Space Telescope as well as complementary observations from the Lowell Discovery Telescope throughout 2024 and 2025. In Section \ref{sec:style} we review the observations and reduction, Section \ref{sec:floats} we present our analysis and results, and in Section \ref{sec:cite} we place our results in context with the current best understanding of the Gateway region and Jupiter system impactors. 

\section{Observations} \label{sec:style}
On 5 December 2024 we obtained eight 370~second Wide Field Camera 3 (WFC3) exposures over two orbits of HST as part of Cycle 32 GO-17795. The observing geometry for the HST observations is presented in Table \ref{tab:obs_geo}. The target was not immediately obvious in the individual frames, and no comae was visually identified. To confirm the target's location within the frames all 8 geometric- and photometric calibrated exposures (\textit{.drc} files), were first shifted in \textit{x} and \textit{y} directions to best align stellar trails and the WCS coordinates for the exposures were corrected. Our first attempt to identify V6 was by creating a mosaic of the images, looking for the V6 to appear as a line of eight point sources in the WCS-corrected frames. The coordinates for V6, as queried from JPL Horizons for version 6 of the orbital solution, were overlaid on the resulting mosaic to search for V6 as it moved relative to the stellar background. V6 was not immediately apparent in this method, which we determined to be sensitive to V$<$26.1. 

\begin{figure}
    \centering
    \includegraphics[width=0.48\linewidth,,trim=2.5cm 1cm 2.5cm 1cm,clip]{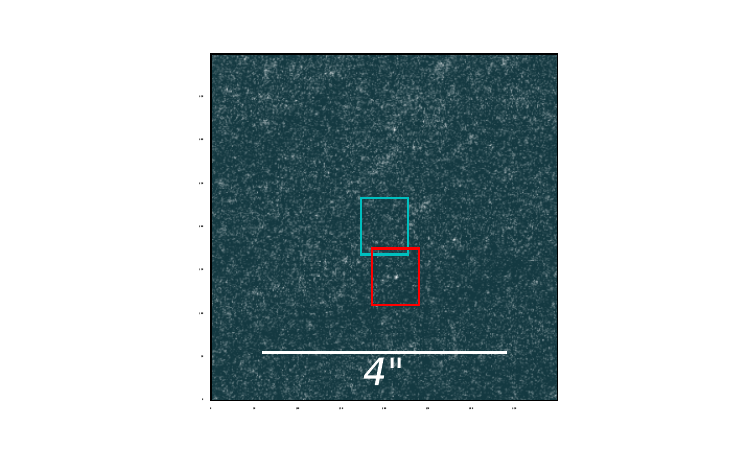}
    \includegraphics[width=0.48\linewidth,trim=2.5cm 1cm 2.5cm 1cm,clip]{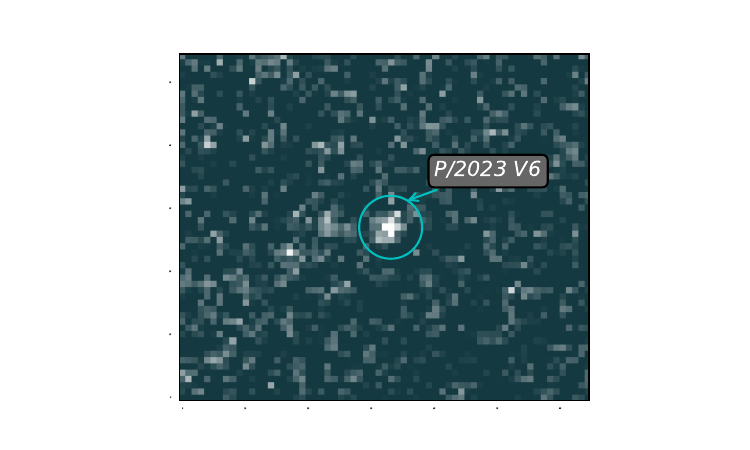}
    \caption{A median-stacked image of all eight 370s science exposures taken by HST WFC3 of P/2023 V6 (PANSTARRS) on December 5, 2024. Each exposure has been shifted in $x$ and $y$ to minimize differences between frames and correct for small differences in the WCS solution. A median stack centered on the JPL predicted location of V6 is shown on the left, with the cyan box highlighting the expected location and the red highlighting the area shown in the image on the right. A 4\farcs~ scale bar, the size of the coma as observed in late 2023, is shown inset in the image on the left for context. In a 5 pixel aperture the signal to noise ratio is 17.  }
    \label{fig:V6_median_stack}
\end{figure}

We then proceeded to stack the images to build signal and probe fainter magnitude limits. To test various combinations of small shifts between images and ensure proper alignment we performed a best fit using the \textit{scipy} \texttt{2dcorrelate} package, allowing us to iteratively test a wide range of shifts in \textit{x} and \textit{y} between images to maximize signal for the only unmoving object within the frame, presumed to be V6. In the resulting stacked image the untrailed nucleus is detected with a signal-to-noise ratio of 17, approximately 0\farcs73 from the calculated JPL Horizons coordinates. Considering that V6 had been active from 2023 through 2024 with a clear dust coma several arcseconds across, our candidate detection's proximity to the predicted position is even more striking considering the \edit{uncertainties in the earlier extended object astrometry}. No other PSF-like sources are present in the full frame image. The median-stacked images are shown in Figure \ref{fig:V6_median_stack}. No coma is immediately apparent in the stacked image, indicating cessation of activity prior to our observations. The odds of detecting a different object at the same tracking rate were briefly considered, but ultimately determined to be far lower than the probability of the object being P/2023 V6. 

\begin{table}[]
    \centering
    \begin{tabular}{c|c|c|c|c|c|c}
        \textbf{Obs.} & \textbf{Date} &\textbf{r$_{h}$ (au)} &\textbf{$\Delta$} & \textbf{RA} &\textbf{Dec} & \textbf{Phase Angle ($^{\circ}$)} \\
        \hline
        \hline
         HST & Dec. 05 2024, 12:30-13:30 UTC& 4.96 & 4.12 & 07 11 28.37-26.38 $\pm$7.580 & +17 36 17.6-18.2  $\pm$ 2.731 & 6.54 \\
         \hline
         LDT & Nov. 22, 2024, 06:22-06:48 UTC& 4.94 & 4.24 & 07 15 25.98-25.71 $\pm$ 6.182 & +17 37 32.3-32.0 $\pm$ 2.451 & 8.70 \\
         LDT & Mar. 31, 2025, 04:00-4:14 UTC& 5.10 & 4.88 & 06 40 31.9-32.12 $\pm$ 7.309 & +19 08 23.3-23.5 $\pm$ 1.921 & 11.16 \\
    \end{tabular}
    \caption{Observing parameters for P/2023 V6. The relevant range of right ascension (RA) and declination (Dec) for the observations and 3$\sigma$ uncertainties are provided in arcseconds. }
    \label{tab:obs_geo}
\end{table}

\section{Results} \label{sec:floats}

\subsection{Activity Trends and Lowell Discovery Telescope Observations}
\begin{figure}[h!]
    \centering
    \includegraphics[width=0.75\textwidth]{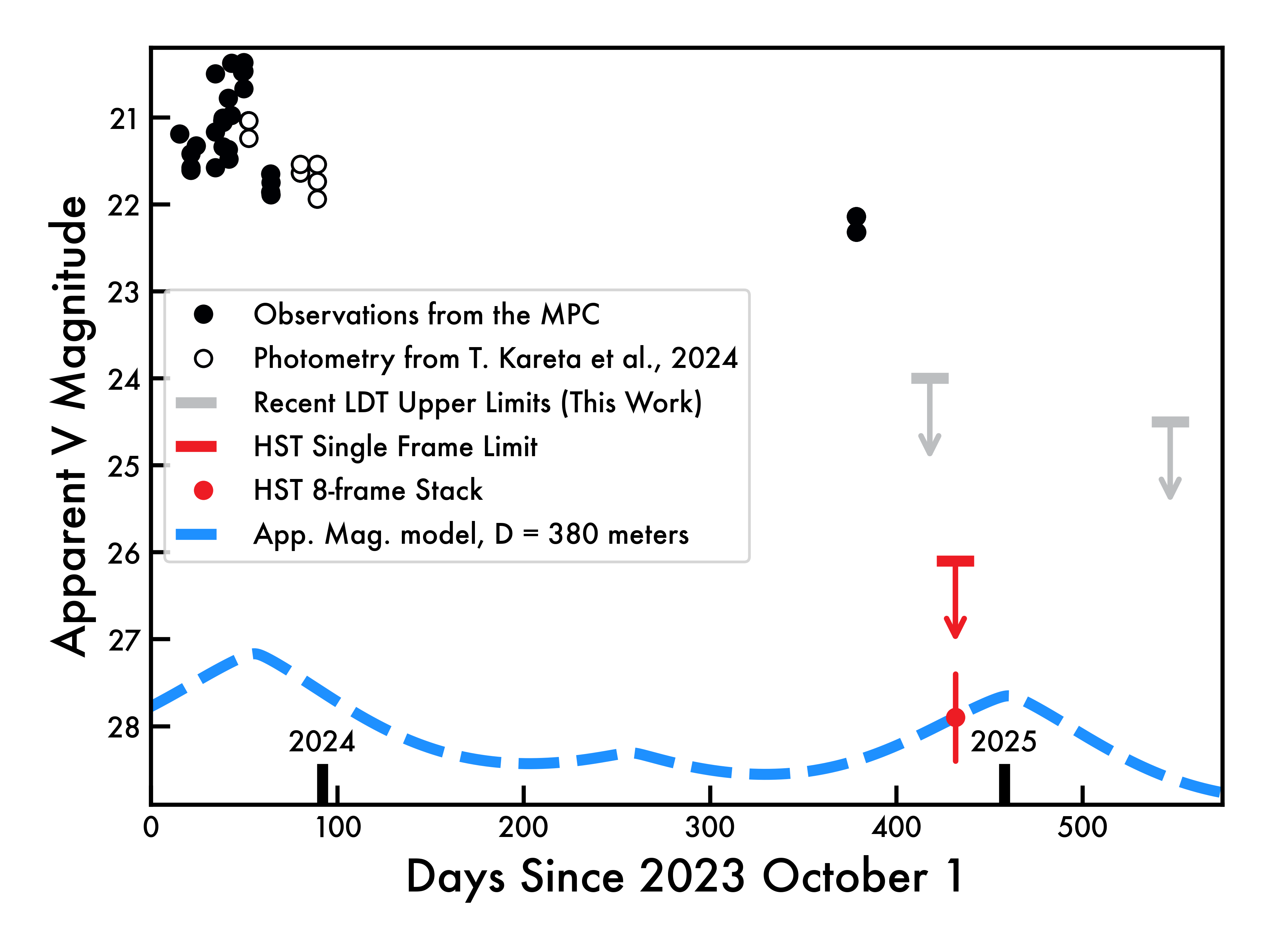}
    
    \caption{A secular lightcurve (e.g., apparent magnitude over time) of P/2023 V6 is shown with data accepted by the Minor Planet Center plotted as black filled circles and the dataset of \citealt{2024ApJ...967L...5K} shown as black unfilled circles. Our recent unsuccessful searches for V6 with the LDT are shown as upper limits in grey while our HST data -- both single-frame upper limits on the brightness of the object and our candidate detection in the stacked HST data at $m_v\approx27.9$ -- are shown in red. A simple model predicting the brightness of the nucleus alone based on the brightness of the object in our stacked HST image is shown as a blue line. All observations were converted to Johnson V for consistency using the colors of the Sun and thus some of the scatter is related to the fact that V6 does not have perfectly Sun-like colors. All observations prior to the HST observations are significantly brighter than the nucleus alone would have been. Activity peaked in the days prior to and around perihelion in November 2023 and then ceased sometime between October and December 2024. The comet then subsequently either fragmented or turned off sometime in October or November of 2024.}
    \label{fig:secular}
\end{figure}

We present an overall summary of the brightness of P/2023 V6 from discovery until March 2025 in Figure \ref{fig:secular}. This includes observations accepted at the Minor Planet Center, the photometric dataset from \citet{2024ApJ...967L...5K}, and new observations from the Lowell Discovery Telescope (LDT) and HST presented in this paper. The object was discovered near magnitude $m_V\approx21.5$ before brightening by approximately another magnitude as it approached perihelion near the end of November 2023. The object then dimmed back to $m_V\sim22$ by the time observations of it ceased in January 2024. The dimming rate was substantially faster than reflected light from a solid nucleus would decrease (an all-nucleus model of the brightness of P/2023 V6 is shown as a blue curve for context in Figure \ref{fig:secular}), indicating that a very significant fraction of the objects brightness was due to its compact coma. The brightening towards perihelion and dimming afterwards, when combined with the lack of outbursts reported photometrically or morphologically \citep{2024ApJ...967L...5K}, suggests that the comet's activity was primarily modulated by changes in its heliocentric distance like LD2 (see, e.g., a discussion in \citet{Kareta2021} as opposed to being outburst-driven like other near-Jupiter comets like 29P/Schwassmann-Wachmann 1 or otherwise discovered in an outburst such that brightness from this discovery epoch could not be reliably used to predict its later behavior. Following the acceptance of our HST program, we proposed for and were granted regularly spaced but short observing runs at the LDT throughout the 2024B and 2025A semesters to monitor the object's activity and perform astrometric measurements as needed.

V6's behavior clearly changes later in 2024. Observations reported to the MPC in October 2024 show the object slightly dimmed over the January to October time frame, the amount of dimming is about what would be expected from changes the object's heliocentric and geocentric distances alone, and thus not an obvious indicator of any change in the object's activity state. We obtained image sequences of P/2023 V6 with the LDT's Large Monolothic Imager (LMI, \citealt{2014SPIE.9147E..2NB}) on 22 November 2024 and 31 March 2025 to contextualize the HST observations described in the next subsection. Other observations were scheduled in addition to these two, but were lost due to weather or otherwise had poor enough conditions (e.g., seeing $>2$") that nondetections were unsurprising. After flatfielding and debiasing the images, standard stars with spectral types similar to that of the Sun were used for photometric calibration of the images within the \textit{PhotometryPipeline} environment \citep{2017A&C....18...47M}. That package was also used for registering the images and for stacking the images. In neither of these image sequences, either in individual images or in a full stack of the images, was the object plausibly near its expected location. We estimated conservative upper limits for how bright V6 could have been in the images based on the brightness of the faintest detected asteroids that coincidentally were in the field of view (a built-in \textit{PhotometryPipeline} tool) and derived upper limits of $m_V > 24$ and $m_V > 24.5$ in November 2024 and March 2025, respectively. This implies that between October and November, the object dimmed by at least $\sim2$ magnitudes and potentially more, and thus that the object had a significant change in the nature of its activity somewhere in that time frame. \edit{This dimming rate is well within the observed rate for active objects at similar distances; the Centaur 29P/Schwassmann-Wachmann 1, which is near 60 km in radius but has a similar thermal state, can dim by 2 magnitudes within 10 days of an outburst \citep{Miles2022,Kareta2025}.}

\subsection{HST Observations and Nucleus Properties}
The brightness of V6 in the HST WFC3 F350LP images is also substantially fainter than expected. Exposure times had been chosen to capture an extended target of total magnitude 21$<$V$<$24, but there is no candidate source with that brightness at the expected position in any individual frame. This results in a single frame lower limit on the magnitude of the object of $m_V > 26.1$. In a stack of all eight frames obtained with WFC3, a single candidate source does show up within $<1$\farcs of the expected position -- a point source at $m_V=28.06\pm0.03$ after converting to the Johnson V filter. The candidate detection can be seen in Figure \ref{fig:V6_median_stack}. Changing the method by which the frames were stacked and aligned did little to change the brightness, location, or appearance of the point source. Given the number of frames combined, the non-sidereal movement of the target, the physical movement of the telescope, and the relatively close match between the expected and measured positions of the comet, we are inclined to think that this detection of V6 is real -- but the faintness of the object is certainly surprising.

If we interpret this as an observation of the nucleus, this corresponds to an absolute magnitude of $H_V=21.49\pm0.03$ if we assume a $0.04$ mag/$^{\circ}$ phase curve typical for cometary nuclei. For a typical cometary albedo of $p_V\approx0.05$, this would make V6's diameter $D = 339_{-5}^{+5}$ meters, almost certainly the smallest known comet at this or any \edit{other} distance. Even if the detection were spurious for some yet-discovered reason and the single frame upper limit of $m_V > 26.1$ were more accurate, this would still restrict the diameter of V6 to below $D < 850$ meters. To put this in a broader context \citet{2017AJ....154...53B} reported a mean diameter for the JFCs of $D\approx1.3$ kilometers -- in either scenario, P/2023 V6 must be smaller than the vast majority of JFCs. \edit{The smallest directly measured JFC nucleui are between 420 and 430 m in diameter, like C/2009 WX51 (D=430$\pm$100 m) and 147P/Kushida–Muramatsu (D=420$\pm$10 m)  \citep[see also 76P/West-Kohoutek-Ikemura and 87P/Bus for similar small objects]{2017AJ....154...53B,Lamy2011,Knight2024} while the more recently characterized dark comet 2003 RM is calculated to be 230 m in radius based on NEOWISE data \citep{Masiero2020,Seligman2024}. There are smaller known comets; the near-Sun comets 322P/SOHO and 323P/SOHO, which have measured nuclei or fragment radii between 20 and 300 m, but are difficult to directly compare with V6, as they are either potentially asteroidal in origin, fragmenting, and have substantially more exotic thermal states \citep{Knight2016,Hui2022}.  We note that these are all substantially smaller in radius than the km-scale mounds on Arrokoth \citep{Stern2023}, suggesting a distinct difference in relevant reservoirs and/or formation mechanism.}

\section{Discussion} \label{sec:cite}

\subsection{What happened to P/2023 V6?}
Sometime between October 13 ($R_H=4.89$ au) and November 22 ($R_H=4.94$ au) of 2024, P/2023 V6 dimmed by at least two magnitudes. By early December 5 ($R_H=4.96$ au) of that year, V6 was about five and a half magnitudes fainter than it had been two months earlier. As seen by Hubble, the comet appeared without any obvious sign of coma or tail -- not only was the object apparently inactive (or very close to it), the inactivity had happened long enough prior to the HST observations that the dust was not detectable in the object's immediate vicinity. This invites an interesting limit on the dust grain size frequency distribution that is dependent on when activity ceased; the closer to the HST observations that the activity declined, the steeper the dust grain distribution must be in order to completely disperse in time, as the smallest grains can be accelerated by the solar wind more rapidly \citep{Burns1979}. $\mathrm{Af\rho}$, a measure of dust abundance that is aperture agnostic, scales linearly with dust mass if the dust properties are assumed to be the same \citep{AHearn1984}. We find that the $\mathrm{Af\rho}$ from the October 2024 MPC observations at V=22.0 have an $\mathrm{Af\rho}$ of $\sim$42 cm, while our observations in December have an $\mathrm{Af\rho}$ of $\leq$0.5, indicating that less than 1$\%$ of the dust remains in the aperture, a clue to the size of particles that could not be accelerated away sufficiently by solar radiation pressure. That sets a constraint on the abundance of large grains \edit{remaining in} the 3,500 km aperture \edit{on 5 December 2024} to be less than 0.9\% of the total mass in October 2024. We can solve for the power law that allows to find an upper limit on the value of $\alpha$ that contains our upper limit mass estimate and confirm our earlier hypothesis that the dust SFD must be steep.

Using the \texttt{sbpy} Python package's \texttt{dynamics} module for generating syndynes and synchrones \citep{Mommert2019} for P/2023 V6 we were able to directly determine the $\beta$ values of particles that remain within the 1.2'' aperture in the 53 days between the observed activity and our HST nucleus detection \citep{Burns1979}. We find that particles with $\beta~<~$0.01, corresponding to particles larger than 22 $\mu$m for cometary like albedo and density, should have remained in the aperture for this duration. In turn, by solving for the size frequency distribution (SFD) $\alpha$ value we can derive a dust mass estimate, which scales linearly with the effective particle radius. We find that that in order for $\leq$1$\%$ of the dust mass to be contained in particles $>$22 $\mu$m in radius the power law of the SFD needs to be \secondedit{$\alpha \geq$ 3.9, consistent with JFCs like 67P \citep[][, and sources therein]{Kelley2008,Marschall2020,Pfeifer2024}}. From this we can derive an \edit{effective} radius of \secondedit{2 $\mu$m when considering particles between 0.1 $\mu$m \citep[the smallest particles observed by the MIDAS instrument at 67P/Churyumov-Gerasimenko, see][]{Kim2023} and 10 cm }\edit{, as demonstrated by Equation 9 of \citet{Ishiguro2016}} and turn to deriving a \edit{lower limit on} total \edit{dispersed} dust coma mass. 

We will analyze this particular aspect more comprehensively in a future publication due to the degeneracies introduced by our limited knowledge of dust speeds and active regions for V6, but for this paper we perform a simple estimate of the dust mass that we were sensitive to. The total cross sectional area of the dust that we had sensitivity to in the full image stack can be written as in \citet{Kim2020}:
\begin{equation}
    C_d = \frac{2.24\times10^{16}\pi}{p_V}10^{0.4[m_{\odot,V}-m_{d,v(1,1,0)}]}
\label{eqn:cx_section}
\end{equation}
where $m_{\odot,F350LP}$=-26.52$\pm$0.03 \citep{Willmer2018}, and $m_{d,v(1,1,0)}$ is the magnitude of the dust signal corrected to 1 au from the Sun, 1 au from the observer, and 0$^{\circ}$ phase angle to match the solar magnitude. We find that in our median stacked image in a 15 pixel aperture ($\approx$1.2'') we were sensitive an extended dust coma of V=27.1, due to the increased background contribution in the larger aperture. This would correspond to an upper limit on $\mathrm{Af\rho}<$0.4 cm and a $m_(d,v(1,1,0)\approx$20.3. Plugging in $m_{d,v(1,1,0)}\approx$20.3 into Equation \ref{eqn:cx_section}, we find $\mathrm{C_d}\approx$13.3 km$^{2}$. 

If we operate under our derived \edit{effective} dust radius of $\bar{a}_d$ = \secondedit{2 $\mu$m} and a density of $\rho_d$ = 2500 kg m$^{-3}$ consistent with the CI and CM chondrites (rather than assuming that cometary bulk porosity persists at these small sizes) \citep{Britt2002,Macke2011}, we can derive an upper limit on the total mass of dust:
\begin{equation}
    M_d= \frac{4}{3}C_d \bar{a}_d \rho_d
\end{equation}
corresponding to a crude estimate of the total dust mass of less than \secondedit{560} kg in the 5 December 2024 WFC3 V6 images. If we compare this to the magnitude reported to the MPC in October 2024, V=22, which corresponds to a total dust mass of \secondedit{89000} kg for the same dust properties used above, \edit{of which} less than 0.9\% remains in the 15 pixel aperture 53 days later. \edit{We note that this derivation is driven by non-detections of the coma, and therefore represents a lower limit on the dust mass lost; we can only constrain the power law for the SFD to be greater than }\secondedit{3.9.}

Due to the faintness of our detection of V6, and thus the possibility (however unlikely) that what we think is V6 in our data is actually correlated noise, let us briefly discuss what a non-detection would imply. While we believe our detection is robust, the vast majority of our conclusions would not change if V6 was not detected in our images due to being even fainter. A non-detection in the eight image stack would imply a size for V6 at least 3-4$\times$ less than our derived value, and thus suggest that the activity seen persisting through October 2024 was effectively catastrophic to the main body. Disruption would naturally produce significant non-gravitational accelerations in any surviving remnants, potentially increasing the uncertainties in the known position of V6. If the ephemeris was sufficiently inaccurate that V6 was not in the HST field of view, the LDT imaging campaign (with a much larger field of view than the relevant uncertainties) supports the dimming of V6 below a magnitude of V$\sim$24 anyways. In any of these scenarios, V6's significant dimming is real -- V6 and LD2 must \edit{have experienced different thermal states and/or initially possessed substantially different volatile reservoirs. Of course, disentangling the primordial properties from the evolutionary ones is difficult, especially without molecular species detections. The rapid dimming still implies an abundance of finer grains in V6 that was not seen in LD2, pointing towards fundamental differences in their activity.}

\textit{Why} the activity would stop in this period \edit{with no major orbital changes, is} at least moderately puzzling. During this period, the comet crossed no major ice lines and its surface temperature should have been stable at the $\%-$level (considering likely albedos and the heliocentric distance involved, at most a two Kelvin difference). Even if the October observations were compromised in some way (e.g., an unnoticed background star contaminated the photometry), comparing the distances of the last observation in the \citet{2024ApJ...967L...5K} dataset ($m_V = 21.6\pm2$, $R_H = 4.56$ au) to the HST data results in a $\sim4.1\%$ temperature decrease (again, about 8-10 Kelvin). While the tendency of water ice to sublimate is indeed changing throughout this temperature range, even $4.56$ au is hardly within most conceptions of where water ice sublimates rapidly. What, then, could cause such a large decrease in the apparent brightness of this comet? We therefore explore the likelihood and implications of the two following scenarios: that P/2023 V6 fragmented and what we see today is some small remnant of a once-larger body, or that V6 suffered no fragmentation event and simply ceased any detectable activity as its available volatile reservoir depleted.

\subsubsection{Did it fragment or fall apart?}
If P/2023 V6 had a partial or nearly-full disintegration or fragmentation event sometime in mid-October, this could explain why it was undetected in later groundbased data (too faint) and why a relatively small nucleus might elude easy detection until the HST observations two months later. A partial disintegration similar to this scenario was suggested to have occurred sometime prior to September 2022 on the active Centaur P/2020 MK4 by \citet{2022RNAAS...6..251R}. Those authors found a very diffuse coma surrounding a $m_V\approx24.9$ point source (fainter than either of our groundbased upper limits on V6 at the same facility) at the expected position of that object following slightly more than a year of no reported photometry at the MPC. While MK4 is likely the only other km-scale Centaur seen to fragment in this way, more minor disruptions in which swarms of large (tens of centimeters and larger) particles are ejected have been seen at 29P/Schwassmann-Wachmann 1 \citep{2019ATel13164....1K} and 174P/Echeclus \citep{2019AJ....158..255K}. Furthermore, the fragment that Echeclus likely released in late 2005 (\citealt{2008PASP..120..393B}, see also alternative interpretations in \citealt{2008A&A...480..543R}) was potentially even larger than our estimated current sizes for V6. In other words, distant JFCs and closer-in Centaurs in similar and even colder orbits than P/2023 V6 seem to disrupt or fragment somewhat frequently, though the absolute occurrence rate -- especially for objects as small as V6 was (or is) -- is hard to estimate. Despite the low temperatures, and thus the smaller amount of energy available to affect physical changes in their nuclei, \edit{these active objects continue to fragment across a range of scales beyond the water ice line}. While our lack of knowledge regarding the frequency of these process does limit our ability to speculate on how the Centaur and JFC SFDs evolve as they approach lower heliocentric distances, more research in this area could clarify the reliability of using these as diagnostics to estimate the initial SFD of the comet-forming portion of the disk.

A primary challenge in invoking a fragmentation event to explain the long term lightcurve of P/2023 V6 is that it is not obvious how an object could fragment, presumably releasing at least some dust and potentially quite a bit, and then have none of that dust or debris be detectable just several weeks later. Radiation pressure does not work very rapidly at such heliocentric distances, and the lack of a prominent anti-Sunward tail in the 2023 observations suggests that the coma at at that time was \textit{not} composed of the fine $\mu{m}-$sized grains capable of rapid removal. If V6 released some dust during this hypothetical event, one might easily imagine it brightened somewhat -- considering that the object was detected just dimmer than the limiting magnitude of typical surveys ($m_V=21.5-22$), the lack of detections by other facilities like the Catalina Sky Survey or PANSTARRS during this timeframe is puzzling. Furthermore, the detection of a point source in our HST data at $m_V=28.06\pm0.03$ speaks strongly to the sensitivity of those observations -- if there was dust nearby, even at a level more diffuse than was seen at P/2020 MK4 \citep{2022RNAAS...6..251R}, we would have been sensitive to it. If the fragmentation had happened earlier (e.g., if the October Gemini observations were not fully reliable) or had happened over a more prolonged period of time \citep[like 73P/Schwassmann-Wachmann 3][e.g., weeks]{Crovisier1996,Boehnhardt2002, Fuse2007}, these constraints become easier to meet. That said, even if invoking a rare and interesting scenario occurring just when observers were not looking is a plausible explanation, other more mundane options ought to be explored.

\subsubsection{Did it just turn off?}
\edit{Our observations could also be explained if P/2023 V6 is a very small comet with a high active fraction that ceased cometary activity at some point between October 13th and November 22nd.} The most natural \edit{comparison,} the other Jupiter Co-Orbital P/2019 LD2 (ATLAS), is also small \citep{2021PSJ.....2...48K} and highly active (see, e.g.,  \citealt{2021PSJ.....2...48K, 2021A&A...650A..79L, 2021AJ....161..116B}). \citet{2024ApJ...967L...5K} proposed that V6 was likely to be larger than LD2 and less active per unit area, but perhaps the opposite is true. 

LD2's $Af\rho$ \citep{AHearn1984} value at $R_H=4.58$ au was $484\pm20$ cm \citep{2021PSJ.....2...48K}, while V6's $Af\rho$ value at $R_H=4.53$ au was just $32\pm2$ cm \citep{2024ApJ...967L...5K} -- a factor of $\approx15$ different. \citet{2021PSJ.....2...48K} constrained the size of LD2's nucleus on the basis of precovery non-detections to be less than $D_{nuc}<2.4$ km assuming a typical $5\%$ albedo for comets. If V6 and LD2 were to be about as active as each other in terms of dust loss per unit area, we would expect that the surface area of LD2 should be $\sim15\times$ larger than V6's and thus that V6's radius should be about a quarter of LD2's. Our HST observations are consistent with a nuclear diameter of just $D_{nuc} = 339_{-5}^{+5}$ meters for V6 (assuming $p_v$=0.05); even smaller than that relation suggests. Even our single HST frame upper limit ($D_{nuc}<850$m) is nearly small enough to meet that limit. In other words, given our HST data, we see no obvious reason to require LD2 and V6 have different activity states or evolutionary histories -- it's just that V6 is the smallest comet detected directly at this heliocentric distance. 

\edit{The detection of an object as small as V6 from current surveys while strongly active has major implications for future surveys. This detection implies that the Vera Rubin Observatory's Legacy Survey of Space and Time (LSST) should essentially be able to discover all active comet co-orbitals of Jupiter and the innermost Centaurs at slightly larger distances (see, e.g., \citealt{2019ApJ...883L..25S})}. This would enable a much more sophisticated understanding of the Jupiter impactor population (see next sub-section) as well as the mechanics of the Centaur-to-JFC transition. \edit{The observed activity evolution of both LD2 and V6 suggests that} major nucleus events are not only possible at these heliocentric distances, but are also capable of producing drastically different size frequency distributions of dust; LD2's dust is characterized by a much shallower power law slope with an average grain radius of 150 $\mu$m \citep{Kareta2021}. This makes sense because LD2 had stronger activity as evidenced by its larger $\mathrm{Af\rho}$ and thus was able to loft larger grains more efficiently compare LD2's 150 $\mu$m average radius to V6's $\leq$\secondedit{2} $\mu$m). \edit{Future observations of active near-Jupiter comets should aim to be directly sensitive to the SFD of the dust, whether by deeper observations in the visible to measure the dust coma morphology or NIR reflectance spectra to characterize the SFD \citep{Kareta2021}.}

One point raised in \cite{Kareta2024} was that if V6 and LD2 were similar in size, that would indicate a lack of detections of activity from objects \edit{with sizes} more typical of JFCs \edit{($d>$2 km)} in similar orbits, which would presumably produce large dust comae that would be easier to detect. One possibility \secondedit{to discuss }is that the co-orbital population of Jupiter is heavily skewed towards debris created from activity and possibly fragmentation, and we may be seeing activation of subsurface volatile reservoirs from impacting debris\edit{; the size distribution of impactors onto Jupiter is heavily weighted towards the 1-10's of meter size range based on bolide measurements, which impact Jupiter 10-100 times per year, with objects the size of V6 expected to impact the giant planet on the 5 to 10 year timescale \citep{Hueso2010,Hueso2013,Hueso2018}. \edit{In this scenario the duration of activity for these objects} would not be driven by changes to surface temperatures, but by the size of the volatile reservoir exposed by the impact. \edit{Consequently,} LD2's activation would be consistent with a larger reservoir being exposed compared to V6, \edit{but} larger objects would not necessarily produce more dust or gas. This same population \secondedit{would} also impact co-orbitals, albeit at much smaller collision frequencies }\secondedit{due to the comparatively minuscule cross-section, so by scaling the cross-sectional ratio of V6 and Jupiter's collisional cross section for short period comets \citep{DellOro2001}, we find on the order of one impact every 100 million to billion years for 1-10 meter sizes \citep{Bottke2023}. The smallest relevant impactors for the co-orbitals would be the $\sim$10 cm-scale, which are capable of exposing subsurface ice within a few meters of the surface and may yield meaningful activity increases. Even the impact frequency for these objects on the coorbitals is still on the order of every hundred thousand years \citep{Bottke2023}, making collision an unlikely activity source.} \edit{Thus there is a clear weakness to this theory, that in order to have multiple impacts on the Jupiter co-orbital population capable of initiating significant activity within a few years the impact probability from the smallest (D$<$10 m) parts of the size frequency distribution would have to be far higher than expected \citep{Bottke2023} \secondedit{, making it an unwieldy solution to explain the frequency of active co-orbitals}. The resolution from the L'LORRI camera onboard the Lucy mission to the Trojan asteroids \secondedit{will} provide unique insight into \secondedit{impacting population size frequency distribution} via observations of the crater size frequency distribution on its \secondedit{Trojan} flyby targets\secondedit{, but to the 10 meter scale \citep{Weaver2023}}. Continued searching for active co-orbitals by ground-based surveys, and especially the Vera Rubin Observatory, will provide the coverage necessary to identify more active objects of 18 $<$m$_V <$ 22. However, constraining the nucleus sizes of these objects in order to further investigate the dust grain size and activity trends will require high angular \secondedit{resolution} observations from space observatories like HST and JWST NIRCam, or thermal infrared observations. }

\subsection{Planetary Defense at Jupiter}
If V6 really is this small and was detected essentially due to its activity alone, this means that current surveys (with limiting magnitudes of $m_V < 21.5$ or so) can detect comets as large as what caused the July 2009 impact scar on Jupiter (see, e.g., \citealt{2010ApJ...715L.155S, 2010ApJ...715L.150H}), estimated at $500-1000$ meters in diameter, so long as they are at least somewhat active. That object was not discovered prior to impact and thus its activity state can only be guessed at, but \citet{2010ApJ...715L.150H} estimated through comparison with its effects on Jupiter to those caused by the impact of the Comet D/1993 F3 (Shoemaker-Levy 9) on Jupiter that it might have been rockier and less volatile rich than SL9 was and thus likely not as active. In any case, even though detecting the largest and rarest impactors might be possible with direct detections of their solid bodies, this detection shows that even weakly active smaller objects can be detected by current surveys in the vicinity of Jupiter and thus that the Vera Rubin Observatory's all-sky survey to fainter brightnesses should do radically better.

SL9's impact onto Jupiter facilitated significant advances in our understanding of both small bodies and of the Jovian atmosphere (see a review in \citealt{2004jpsm.book..159H}). While pre-impact detection of the smallest impactors is obviously going to be better pursued with non-telescopic techniques (e.g., in-situ data, such as \citet{2021GeoRL..4891797G} which saw a potential impact in \textit{Juno} data that might have come from a 1-4 meter object), any increase in our ability to find and characterize impactors ahead of time and then prepare follow-up observations to see how Jupiter responds could be tremendously useful for the small body and giant planet communities. 

\subsection{Implications for the Jovian Irregular Satellites}
Activity on \edit{objects of V6's size} in Jupiter's vicinity also prompts us to question how prolonged (like LD2) or sporadic/limited (like V6) activity can influence the orbital evolution of these co-orbitals. Two recent results have highlighted the importance of considering \edit{non-gravitational effects} even for weakly active and distant comets: the discovery of dark comets in the inner Solar System \citep{Seligman2024} which display significant non-gravitational drifts in their orbits without detected dust comae and the realization that many Oort Cloud comets clearly show non-gravitational accelerations well beyond Jupiter \citep{2023A&A...678A.113K}. Given that V6 and LD2 are both small comets and at similar heliocentric distances, we posit that similar non-gravitational forces are likely important for both of these end-member co-orbitals. If that hypothesis is true, then those non-gravitational forces \edit{in the radial and transverse directions would increase the parameter space} for objects that are on orbits likely to evolve into Jupiter co-orbitals \edit{to become temporary captures}, and may increase the likelihood of direct capture into the Jovian system. \edit{Testing such a hypothesis would require identifying a Centaur or JFC with a probability of becoming a temporary co-orbital in the near future, and obtaining accurate astrometry over the preceding period in order to determine the magnitude of any non-gravitational forces. }

If anything, the production of massive amounts of dust by these active objects so near to Jupiter's Hill sphere would provide a current source of likely carbonaceous chondrite dust to blanket the Jovian system. However, recent JWST \edit{surface reflectance spectroscopy} of the irregular satellites appear to show similarity to dead cometary nuclei \citep{Sharkey2025}; however further work is needed to determine how we can distinguish between recent emplacement of a dead comet and the previously proposed theory of early emplacement and gradual evolution \edit{\citep{Nesvorny2007,Nesvorny2014}}. The presence of ammoniated phyllosilicates on Himalia and Elara, the largest of the irregular satellites, is a clue to some of the thermal history of those objects \edit{\citep{Sharkey2025}}. Whether that makes it reasonable for comparison to the spectrum of 67P/Churyumov-Gerasimenko as an analog or as evidence of evolutionary degeneracy is part of future efforts. 




\section{Conclusion}
In observations of C/2023 V6 taken with WFC3 instrument onboard HST we found no signs of activity, and directly detect the nucleus at a magnitude of 28.06$\pm$0.03, corresponding to $H_V$=21.49$\pm$0.03, or a diameter of 339 meters for a cometary albedo of 0.05. The lack of activity, small nucleus size, and status as a Jupiter co-orbital object raises many questions regarding the initiation mechanism, the source of activity, and why the activity ceased not long after the observations in October 2024. We show that less than \secondedit{560} kg of dust could reasonably be within our 3500 km aperture, and that based on the rapid dissipation of dust at nearly 5~au between October and December 2024 that the size frequency distribution of the dust present in the October 2024 observations must have had a power law slope with a \edit{lower limit of $\alpha \secondedit{ \geq 3.9}$}, \secondedit{similar to JFCs}. Such a slope could be indicative of weak localized activity that was unable lift larger grains from the surface, a sign that whatever volatile reservoir had been tapped in 2023 was running dry.

Given that both LD2 and V6 seem to have activated while in relatively stable orbits with little changes to thermal environments we hypothesize that their activity may be initiated by impacts from meter-scale fragments of previous visitors to the co-orbital region. This mechanism would also explain why we have not seen more large active co-orbitals, which presumably would have had proportionally larger dust production rates. With LSST coming online and sensitivity to these V$\approx$22 active objects more uniform, we are on the cusp of being able to perform similar studies to examine the bare nuclei of these fresh Centaurs on a statistically significant level, provided we can obtain consistent astrometry to constrain observational uncertainties to obtain similar deep imaging with HST WFC3 in the future. By doing so we may be able to find potential Jupiter impactors years before their close, and potentially final, encounters with the gas giant. 

\begin{acknowledgments}
Based on observations with the NASA/ESA/CSA Hubble Space Telescope obtained at the Space Telescope Science Institute, which is operated by the Association of Universities for Research in Astronomy. The authors extend their sincere thanks to Carol Rodriguez for helping schedule and execute these challenging observations. Incorporated, under NASA contract NAS5-26555. All authors acknowledge support by HST program number GO-17795 (PI J. Noonan), which was provided through a grant from the STScI under NASA contract NAS5-26555. The HST data presented in this article were obtained from the Mikulski Archive for Space Telescopes (MAST) at the Space Telescope Science Institute. The specific observations analyzed can be accessed via \dataset[doi:  We10.17909/vff8-3p23]{https://doi.org/10.17909/vff8-3p23}. We would also like to acknowledge the efforts of our two reviewers, who provided insightful feedback that greatly improved this manuscript. 
\end{acknowledgments}

%

\vspace{5mm}
\facilities{HST(WFC3)}


\software{astropy \citep{astropy2013,astropy2018,astropy2022}, sbpy \citep{Mommert2019}
}




\end{document}